\documentclass[10pt, conference]{IEEEtran}

\usepackage{algorithm,multirow}
\usepackage[noend]{algorithmic}
\usepackage{graphicx}
\usepackage{comment}
\usepackage{multirow}
\usepackage{amssymb,amsthm,mathrsfs}
\usepackage{amsfonts,epsf,graphicx,times,amsmath,multirow,url}
\usepackage{float}
\usepackage{flexisym}
\usepackage[capitalise]{cleveref}
\usepackage{float}
\usepackage{verbatim}

\usepackage{amsthm}
\usepackage{subcaption}
\usepackage{color}

\usepackage{footnote}
\usepackage{footmisc}
\makesavenoteenv{tabular}
\makesavenoteenv{table}
\graphicspath{ {./figs/} }

\usepackage[numbers]{natbib}

\title{A Hybrid System for Learning Classical Data in Quantum States}

\author{
\IEEEauthorblockN{
Samuel A Stein\IEEEauthorrefmark{1}\IEEEauthorrefmark{2},
Ryan L'Abbate\IEEEauthorrefmark{1},
Wenrui Mu \IEEEauthorrefmark{1},
Yue Liu\IEEEauthorrefmark{1},
Betis Baheri\IEEEauthorrefmark{3},\\
Ying Mao\IEEEauthorrefmark{1},
Guan Qiang\IEEEauthorrefmark{3},
Ang Li\IEEEauthorrefmark{2}
and Bo Fang\IEEEauthorrefmark{2} \\
}
\IEEEauthorblockA{\IEEEauthorrefmark{1} Department of Computer and Information Science,
Fordham University, \\ Email: \{sstein17, rlabbate, wmu2, yliu720, and ymao41\}@fordham.edu}
\IEEEauthorblockA{\IEEEauthorrefmark{2} Pacific Northwest National Laboratory, 
\IEEEauthorrefmark{2} Email: \{samuel.stein, ang.li, bo.fang\}@pnnl.gov}
\IEEEauthorblockA{\IEEEauthorrefmark{3}Department of Computer Science, Kent State University, 
\IEEEauthorrefmark{3} Email: \{bbaheri, qguan\}@kent.edu}
}

\begin{document}
\maketitle
  
\begin{abstract}
Deep neural network powered artificial intelligence has rapidly changed our daily life with various applications. However, as one of the essential steps of deep neural networks, training a heavily-weighted network requires a tremendous amount of computing resources. Especially in the post Moore's Law era, the limit of semiconductor fabrication technology has restricted the development of learning algorithms to cope with the increasing high intensity training data. Meanwhile, quantum computing has demonstrated its significant potential in terms of speeding up the traditionally compute-intensive workloads. For example, Google illustrated quantum supremacy by completing a sampling calculation task in 200 seconds, which is otherwise impracticable on the world's largest supercomputers. To this end, quantum-based learning has become an area of interest, with the potential of a quantum speedup. In this paper, we propose GenQu, a hybrid and general-purpose quantum framework for learning classical data through quantum states. We evaluate GenQu with real datasets and conduct experiments on both simulations and real quantum computer IBM-Q. Our evaluation demonstrates that, compared with classical solutions, the proposed models running on GenQu framework achieve similar accuracy with a much smaller number of qubits, while significantly reducing the parameter size by up to 95.86\% and converging speedup by 33.33\% faster. 

\end{abstract}

\section{Introduction}
In the past decade, machine learning and artificial intelligence-powered applications have dramatically changed our daily life. Many novel algorithms and models achieve widespread practical successes in a variety of domains such as autonomous cars, healthcare, manufacturing, etc. Despite the wide adoption of ML models, training the machine learning models such as DNNs requires a tremendous amount of computing resources to tune millions of hyper-parameters.
Especially in the post Moore's Law era, the limit of semiconductor fabrication technology cannot satisfy the the rapidly increasing data volume needed for training, which restricts the development of this field~\cite{thompson2020computational}.



Encouraged by the recent demonstration of quantum supremacy~\cite{arute2019quantum}, researchers are searching for a transition from the classical learning to the quantum learning, with the promise of providing a quantum speedup over the classical learning. 
The current state of quantum-based learning inspires alternative architectures to classical learning`s sub-fields, such as Deep Learning (DL) or Support Vector Machine (SVM) \cite{stein2021quclassi,beer2020training, potok2018study,levine2019quantum, stein2020qugan}, where the quantum algorithm provides improvements over their classical counterparts. 
For example, there are quite a number of adoptions of quantum learning algorithms in domains of expectation maximization solving (QEM) \cite{kerenidis2019quantum} that speeds up the kernel methods to sub-linear time \cite{li2019sublinear}, Quantum-SVM~\cite{ding2019quantum}, and NLP \cite{panahi2019word2ket}.
The use of quantum systems to train deep learning models is rather developed with a multitude of approaches for creating and mimicking aspects of classical deep learning systems \cite{verdon2019learning,beer2020training,chen2019variational,kerenidis2019quantum}. 
For example, Tensorflow Quantum~\cite{broughton2020tensorflow} provides a quantum ML/DL library for rapid prototyping of hybrid quantum-classical models. However, the applications cannot be executed directly on near-term quantum devices, such as IBM-Q platform. 
These quantum systems, in the literature, face several challenges: (i)
such systems are held back by the low qubit count of current quantum computers. (ii) learning in a quantum computer becomes even more difficult due to the lack of efficient classical-to-quantum data encoding methodology~\cite{zoufal2019quantum, cortese2019system}. (iii) most of the existing studies are based on purely theoretical analysis or simulations, lacking practical usability on near-term quantum devices~\cite{preskill2018quantum}.

More importantly, the above challenges would persist even as the number of qubits supported in quantum machines increases significantly: when the number of qubits in the quantum system increases,
the computational complexity grows exponentially~\cite{kaye2007introduction}, which quickly leads to tasks that become completely infeasible for simulation and near-term quantum computers. Therefore, discovering the representative power of qubits in quantum-based learning systems is extremely important, as not only does it allow near-term devices to tackle more complex learning problems, but it also eases the complexity of the quantum state exponentially. However, current research on the topic of low-qubit counts of current quantum machines is rather sparse
Within this domain, the learning potential of qubits are under-investigated. 

In this paper, we propose {\bf GenQu}, a general-purpose quantum-classic hybrid framework for learning classical data in quantum states.
We demonstrate the power of qubits in machine learning by approaching the encoding of data onto a single qubit and accomplish tasks that are impossible for comparative data streams on classical machines, which addressing the challenges (i) and (ii). Enabled by {\bf GenQU}, we develop a deep neural network architecture for classification problems with only 2 qubits, and a quantum generative architecture for learning distributions with only 1 qubit, and, additionally,
We evaluate {\bf GenQU} with intensive experiments on both IBM-Q real quantum computers and simulators (addressing the challenge (iii)). Our major contributions include the following:

\begin{itemize}
  \item We propose GenQu, a hybrid and general-purpose quantum framework that works with near-term quantum computers and has the potential to fit in various learning models with a  low-qubit count.
  
  \item We propose three different quantum-based learning models based on GenQu to demonstrate the potential of learning data in quantum state. 
  
  \item We show through experiments on both simulators and IBM-Q real quantum computers that models in GenQu are able to reduce parameters by up to 95.86\% but still achieves similar accuracy in classification with 
Principal Component Analysis (PCA)\cite{hoffmann2007kernel} MNIST dataset, and converge up to 33.33\% faster than traditional neural networks.
\end{itemize}


\section{Preliminaries}
\subsection{The Quantum Bit (Qubit)}
Quantum computers operate on a fundamentally different set of physics when compared to classical computers. Classical computers operate on binary digits (bits), represented by a 1 or a 0. Quantum computers however, operate on quantum bits (qubits). Qubits can represent a 1 or a 0, or a probabilistic mixture of both 1 and 0 simultaneously, namely superposition. When discussing a quantum framework, we make use of the $\langle bra|$ and $|ket \rangle$ notation, where a $\langle bra|$ indicates a horizontal state vector ($1\times n$) and $|ket\rangle$ indicates a vertical state vector ($n\times1$). A qubit represented as a probabilistic combination of both a $|1\rangle$ and $|0\rangle$. and therefore is described as a linear combination between of $|0\rangle$  and $|1\rangle$. This combination is described in Equation \ref{eq:base_equations-1}.
\begin{equation}
|\Psi\rangle=\alpha|0\rangle+\beta|1\rangle
\text{ , }|\Psi\rangle=\left[\begin{array}{l}
\alpha \\
\beta
\end{array}\right] \text{ , }
\label{eq:base_equations-1}
\end{equation}

\begin{figure}[ht]
    \centering
    \includegraphics[width=0.4\linewidth]{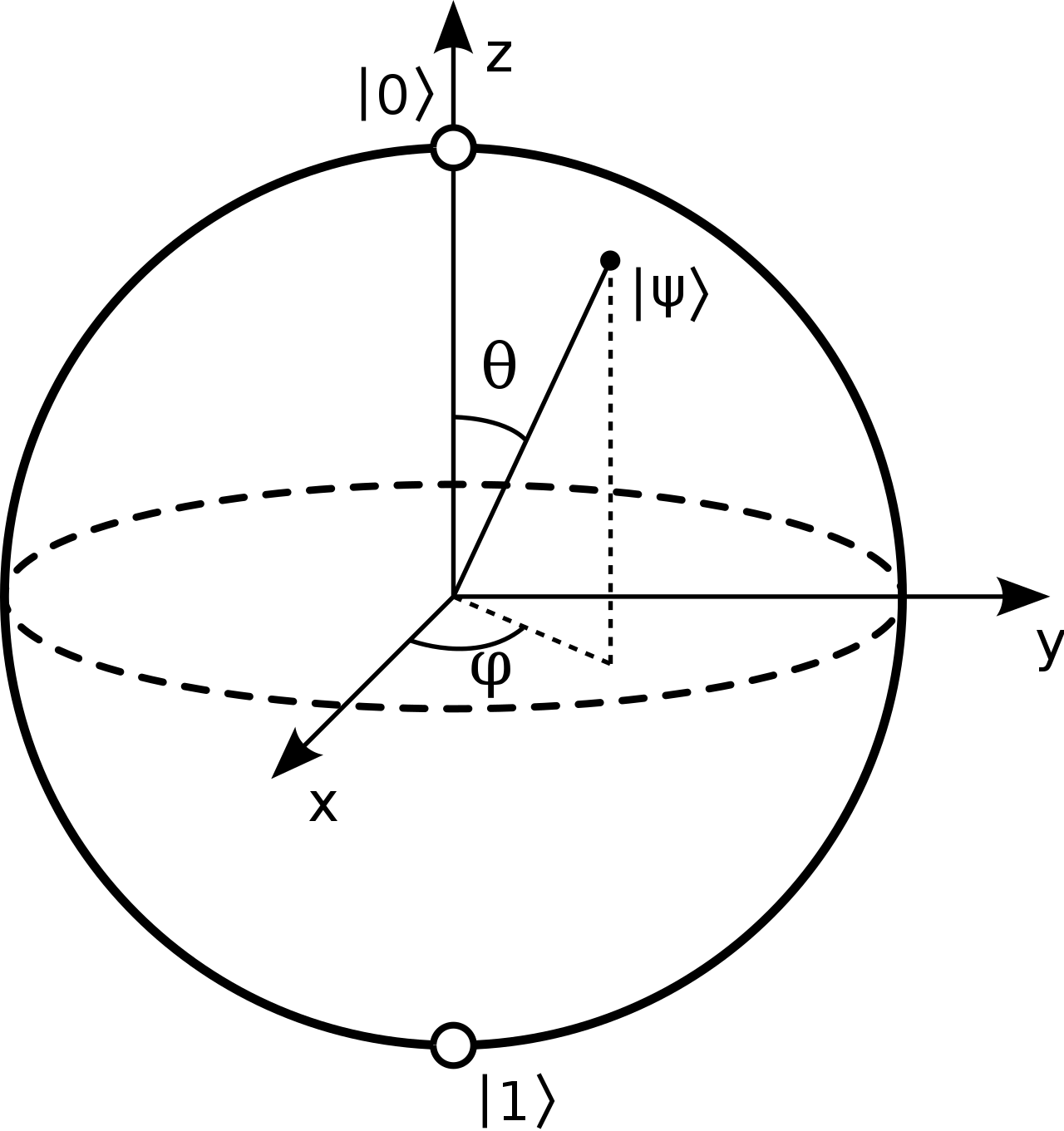}
    \caption{Bloch Sphere}
    \label{fig:bloch_sphere}
\end{figure}

In Equation \ref{eq:base_equations-1}, the state is described as a combination of both $|0\rangle = \left[\begin{smallmatrix}1 \\ 0\end{smallmatrix}\right]$ and $|1\rangle = \left[\begin{smallmatrix}0 \\ 1\end{smallmatrix}\right]$. The values of $\alpha$ and $\beta$ are the probability coefficients and what encode information regarding this qubit's state. Although qubits can exist in both $|1\rangle$ and $|0\rangle$ at the same time, when they are measured for a definite output, they collapse to one of two possible values, where in the case above those values are $|0\rangle$ or $|1\rangle$. The coefficients, $\alpha$ and $\beta$, indicate the square root of the probability that the qubit measures as a $|1\rangle$ or a $|0\rangle$. The definite states we are measuring the qubit against are based on how we measure the qubit, measuring as one of two possible measurements. These two possible measurements are two orthogonal eigen-vectors, and can be in any 3-Dimensional direction. This is best visualized and understood by the Bloch Sphere representation of a qubit, as illustrated in Figure \ref{fig:bloch_sphere}.

A qubit can be represented by the unit Bloch Sphere visualized in Figure \ref{fig:bloch_sphere}. In the case of $|0\rangle$ and $|1\rangle$, we are measuring across the Z axis. 

\subsection{Quantum Data Manipulation} 
Qubits are manipulated through quantum gates, which in turn manipulates the overall quantum state. These gates can allow for manipulation over the state vector, which describes the state of any number of qubits. We introduce the few gates that we make use of in this paper in Equations 3, 4, 5, and 6.
\begin{equation}
R_{Y}(\theta)=\left[\begin{array}{cc}
\cos \left(\frac{\theta}{2}\right) & -\sin \left(\frac{\theta}{2}\right) \\
\sin \left(\frac{\theta}{2}\right) & \cos \left(\frac{\theta}{2}\right)
\end{array}\right]
\label{eq:squ-3}
\end{equation}

\begin{equation}
R_{Z}(\theta)=\left[\begin{array}{cc}
e^{\frac{-i\theta}{2}} & 0 \\
0&e^{\frac{-i\theta}{2}}
\end{array}\right]
\label{eq:squ-4}
\end{equation}

\begin{equation} 
\centering
CR_Y(\theta)=\left[\begin{array}{cccc} 
1 & 0 & 0 & 0 \\
0 & 1 & 0 & 0 \\
0 & 0 & \cos(\frac{\theta}{2}) & -\sin(\frac{\theta}{2}) \\
0 & 0 & \sin(\frac{\theta}{2}) & \cos(\frac{\theta}{2}) 
\end{array}\right]
\label{eq:gates-5}
\end{equation}

\begin{equation} 
\centering
CR_Z(\theta)=\left[\begin{array}{cccc} 
1 & 0 & 0 & 0 \\
0 & 1 & 0 & 0 \\
0 & 0 & e^{\frac{i\theta}{2}} & 0 \\
0 & 0 & 0 & e^{\frac{i\theta}{2}} 
\end{array}\right]
\label{eq:gates-6}
\end{equation}

The gates above accomplish specific state manipulation. Equations 3 and 4 represent single qubit rotations around the Y axis and Z axis respectively. These two gates allow for a single qubit to be manipulated to any position on a Bloch sphere's surface, from any starting point on aforementioned sphere. Equations 5 and 6 entangle two qubits using controlled qubit rotations around the Y axis and Z axis respectively. Controlled rotations allow for a qubit's state to be manipulated based on whether a control qubit measures as a $|1\rangle$, therefore entangling the target and control qubits' states. These two styles of gates, single qubit rotations and controlled qubit rotations, allow for complete manipulation of quantum states, be it 1 or more qubits.

\begin{figure}[ht]
    \centering
    \begin{subfigure}{0.48\linewidth}
        \centering
        \includegraphics[width=1\linewidth]{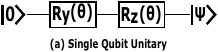}
    \end{subfigure}
    \begin{subfigure}{0.48\linewidth}
        \centering
        \includegraphics[width=1\linewidth]{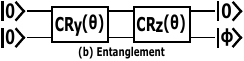}
    \end{subfigure}
    \caption{Gate Deep Learning Gate Design}
    \label{fig:layered_gates}
\end{figure}

\subsection{Quantum Deep Learning}
Quantum Deep Learning is a relatively new approach to Quantum Machine Learning that takes quantum circuits and applies variational quantum algorithm techniques and learning methods similar to those of classical neural networks ~\cite{chen2019variational,garg2020advances,beer2020training}. 
In traditional deep learning, layers are often used, where a layer is some large transformation function that takes in a set of inputs, and transforms them to a set of outputs. These functions are connected and typically trained through the use of back 
propagation~\cite{goodfellow2016deep,chen2019variational}. Similar to how classical deep learning works, the way this data flows through time is up to the practitioner, who chooses and designs their network according to their needs. Quantum deep learning is approached through the use of layering gates sequentially. For this paper, our layers are comprised of the gates in Equations 3 through 6. Similar to how deep learning is parameterized by connection weights, these gates are parameterized through rotations ($\theta$). At the end of the quantum circuit, a loss function is described, and the quantum networks parameters $\theta$ are updated iteratively such that the circuit's loss is optimized \cite{beer2020training,crooks2019gradients}. 


\section{GenQu Framework and learning models}
\subsection{GenQu Framework}

\begin{figure}[ht]
    \centering
    \includegraphics[width=0.8\linewidth]{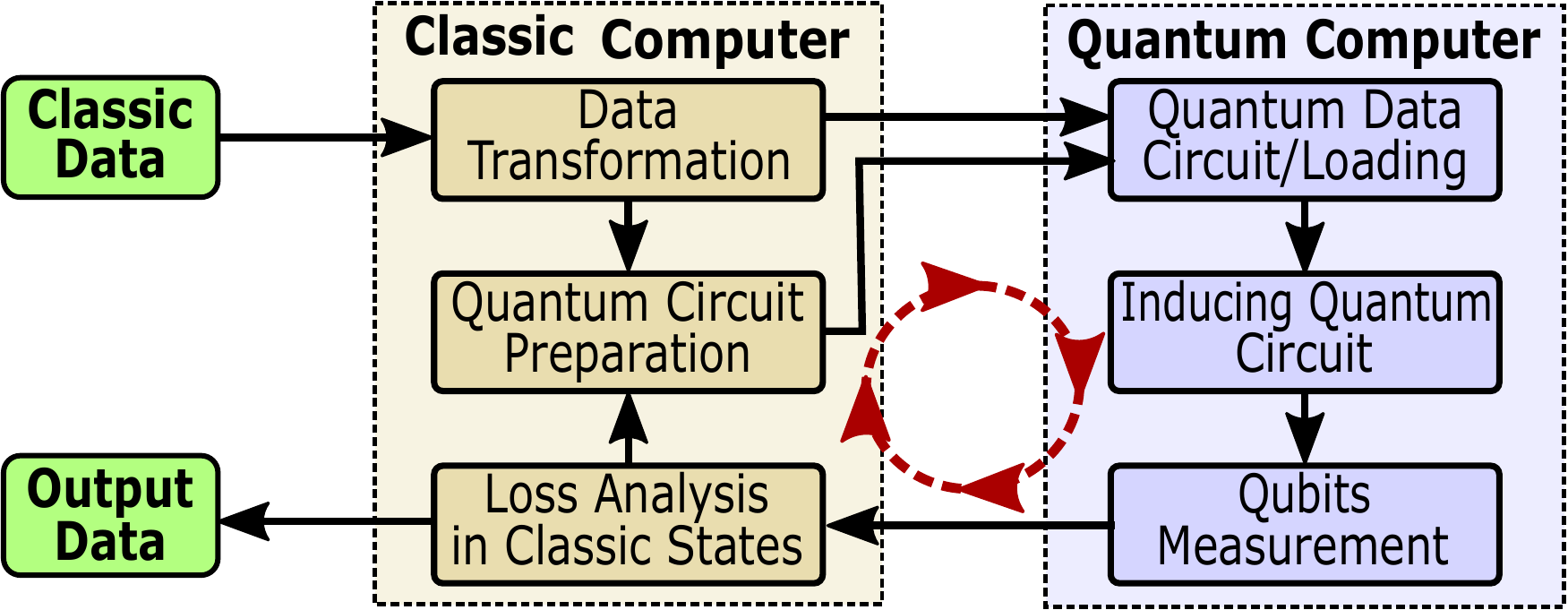}
    \caption{GenQu: A Hybrid Framework}
    \label{fig:system}
\end{figure}
Our proposed GenQu framework is illustrated in Figure \ref{fig:system}. Before any operation of the framework is performed, the data must be transformed from classical to quantum states. This is done by transforming classical data into applicable quantum rotations, and is described under section Section~\ref{qubitization}. Following this transformation, the rotations are loaded onto a quantum computer. The quantum circuit preparation section is where the circuit relating to a specific machine learning algorithm is designed. For example, this is where a deep neural network or a convolutional neural networks architecture would be set up, initialized, and prepared. This circuit is loaded onto the quantum computer after the quantum data loading section. 

Once the circuit is set up, it can be induced. Inducing the quantum circuit results in the quantum state transformation of the input data over the quantum machine learning model. From here, if the output of the model was a quantum state, one could end here and feed it to another quantum algorithm. However, in the case of updating learn-able parameters, the relevant qubits need to be measured. We feed the qubits measurements to a loss analysis section, where we update our parameters accordingly. Once the parameters have been updated, we repeat this process of circuit loading, circuit inducing, and measurement, updating parameters until a desired loss of the network is attained or a predefined number of epochs have achieved. 

\subsection{Data Qubitization}
\label{qubitization}
Prior to discussing our methods of illustrating the learning power of qubits, we introduce our approach to encoding classical data into quantum states. We encode two dimensions of data per qubit, by the simple two step process outlined in Equations \ref{eqn:dim1} and \ref{eqn:dim2}
\begin{equation}
    x_1 \xrightarrow[|\phi\rangle]{\text{Encoded onto}} = RY(2sin^{-1}(\sqrt{x_1}))
\label{eqn:dim1}
\end{equation}
\begin{equation}
     x_2 \xrightarrow[|\phi\rangle]{\text{Encoded onto}} = RZ(2sin^{-1}(\sqrt{x_2}))
\label{eqn:dim2}
\end{equation}
 The value of $x_1$ is encoded along the Z-axis, followed by $x_2$ being encoded along the Y-axis. For these rotations to work, the data along each dimension must be normalized to be in the range of (0,1). This reduction in qubit count is pertinent for the case of quantum machine learning as the state space vector of a quantum system is of tensor rank $2^n$ values, and therefore halving the qubit count provides a $2^{\frac{n}{2}}$ reduction in state space.

\subsection{Single Qubit Kernelized Classification}
When tackling a classification task on a quantum system, we want to encode our data such that the probability of measuring a $|1\rangle$ is comparative with the probability of classifying a data point as $Class$=1. Therefore, in the case of classical data sets, we can wrap 2 dimensions of data around a qubit such that we maximize the ability of the qubit to classify the data. In the case of the circles data set, a data set comprised of points non-linearly separable, we can wrap the qubit with data points such that the rotation around the Z axis is correlated to the distance from the circle center. This is visualized in Figure \ref{fig:circles_classical}. This encoding accomplishes on one qubit the encapsulation of 2 dimensions of non-linearly separable data, whilst accomplishing a separation task. For this to be done, two parameters per qubits are used to transfer between Classical data to Quantum state. These parameters are the rotations around the Y axis, proceeded by a rotation around the Z axis. 

\begin{figure}[ht]
    \begin{subfigure}{0.59\linewidth}
    \centering
    \includegraphics[width=1\linewidth]{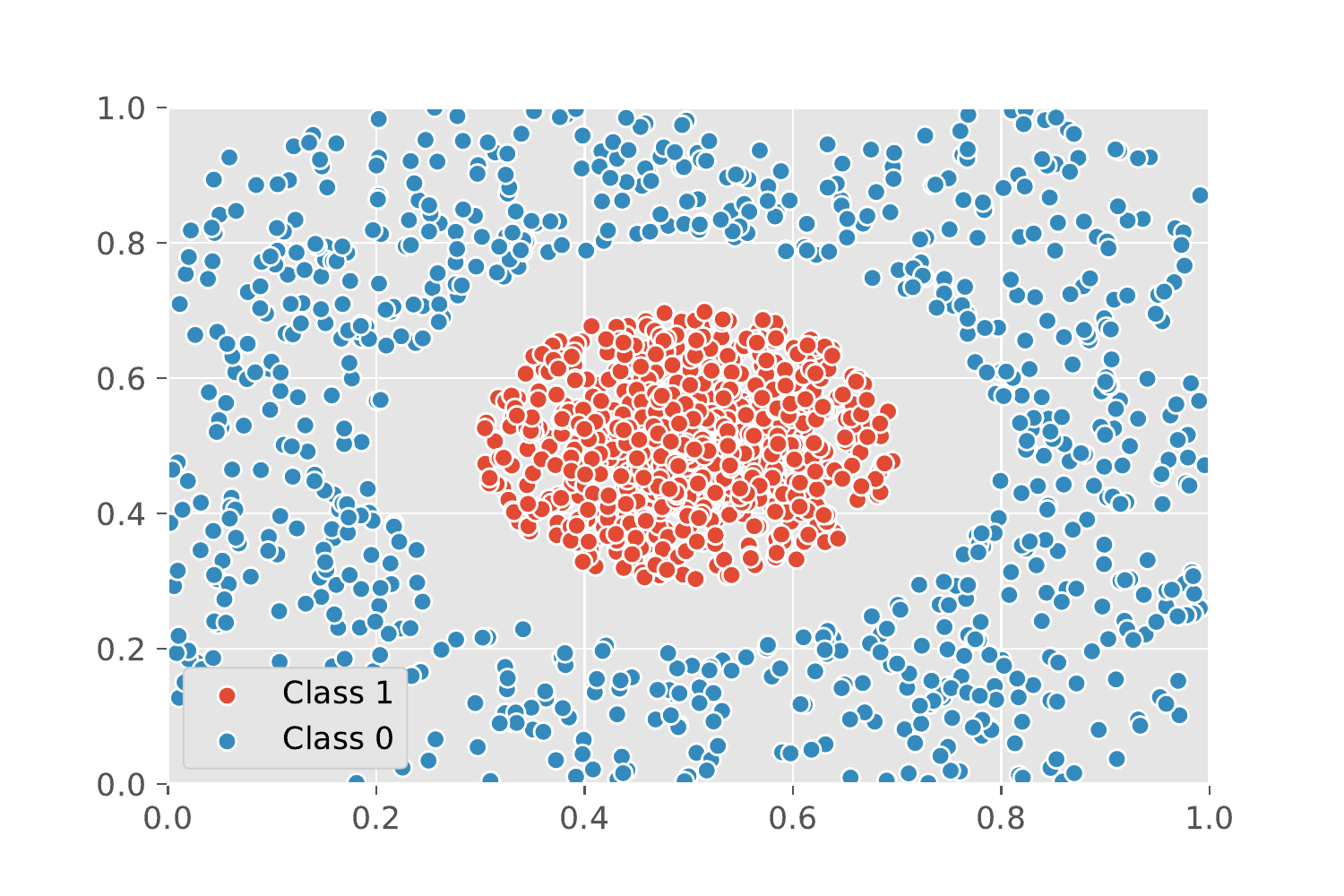}
    \caption{Bit representation)}
    \label{fig:circles_classical}
    \end{subfigure}
\begin{subfigure}{0.39\linewidth}
    \centering
    \includegraphics[width=1\linewidth]{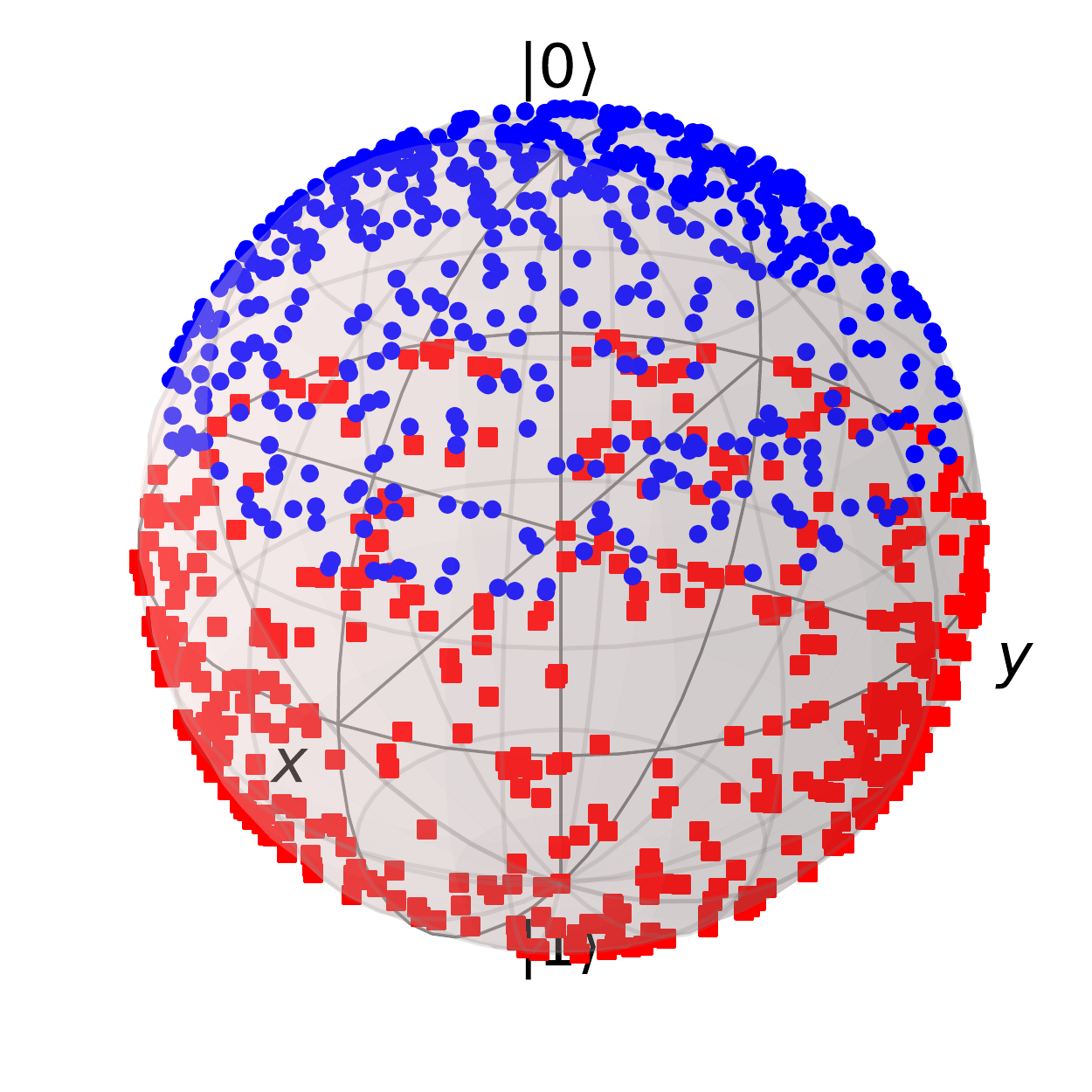}
    \caption{Qubit representation}
    \label{fig:circles_classical}
\end{subfigure}
\caption{Cicle dataset with bit and qubit representations}
\end{figure} 


Translating the classification problem of the data outlined in Figure \ref{fig:circles_classical} to GenQu framework, we follow the following approach. We begin by translating our data into rotations according to the functions outlined in Section~\ref{qubitization}, however the values encoded are vector distances from the circle center. This can be considered both the Quantum Circuit Preparation and Data Transformation components, as the expected measurement under Qubits Measurement is equivalent to the classification of a data point. There is no updating of parameters in this case, therefore we do not iterate and update circuit parameters.

\subsection{Quantum Deep Learning Architecture}
In this paper we make use of the data qubitization techniques outlined above, along side current quantum machine learning techniques to enable highly performent quantum deep learning. Through encoding 2 dimensions of data per qubit, the number of {\em neurons} in our network input is half the dimensionality of the data set. Quantum deep learning layers comprised of single qubit operations are namely called single qubit unitary layers. In these layers each qubit has a RY and RZ gate appended, thereby adding $2n$ parameters, where n iWs the number of qubits. Another type of layer consists of operations acting on two qubits per gate, where operations are control operations (CRY or CRZ). These are namely Entanglement layers. Entanglement layers entangle all qubits by some learnable amount, performing CRY and CRZ gates on qubits $i$ and $i+1$ until there are no qubits left to pair. Entanglement layers require $2(n-1)$ parameters. These gates are visualized in Figure \ref{fig:layered_gates}. Through the use of the entanglement layer, we can reduce and grow the number of qubits at any time point across a circuit dynamically. In our case of illustrating binary classification, we make use of the entanglement layer to pool down data from the other qubits onto one qubit, which is measured and used as the classification qubit. The probability of the qubit measuring $|1\rangle$ is thought of as the probability of labelling the quantum data that was fed to the circuit as $Class =1$ similar to how a single output neuron operates of activation function Sigmoid operates in classical neural networks. 

Fitting this Quantum Deep Learning model to our GenQu framework, we begin by translating our data into rotations described under Section~\ref{qubitization}. From here, the practitioner can describe their full quantum deep learning architecture and initialize the parameters. The data is loaded onto a quantum computer in series, with the quantum data loading circuit being appended with the quantum deep learning architecture. The quantum circuit is induced and the classification qubit measured. We feed this result back to a classical computer, calculate our loss and update our parameters accordingly. This is repeated until convergence occurs or sufficient accuracy is attained.

\subsection{Quantum Generative Nature}
Another powerful use of qubits is in the representation of data. Through using trainable circuits as discussed above, we can measure two values from one qubit. Therefore, a quantum deep neural network can be trained to mimic some input data by defining some loss function such as the Mean Squared Error, and generate new samples that are close to what the qubit was trained on. This is similar to Generative Adversarial Networks~\cite{goodfellow2014generative}, but does not take any noise as an input, nor does it require two networks to be used. We do not claim that ours is better, however it is one of the side effects of qubits being used to represent data, and quantum deep learning models. Therefore, we illustrate through the use of quantum deep learning how a quantum deep learning architecture with a tuned loss function can generate data similar to that of the data it was fed, and at a generative diversity significantly greater that is unattainable using similar architectures within its classical counterparts.

Translating a quantum generative state to GenQu framework, we repeat the steps outlined in the Quantum Deep Learning architecture above. However, the only change would be the loss function such that the quantum state instead of a loss function such as cross entropy, could be mean squared error or some other applicable loss function. Furthermore, no data loading for input is necessary, and instead are just loaded as qubits in the state of $|0\rangle$. When generating data, the qubits are measured and sent to the Output Data stream. 

\section{Results}
We implement GenQu with IBM Qiskit and Tensorflow Quantum. It is evaluated with the above mentioned three applications, kernelized classification, quantum deep learning, and quantum generative nature. We evaluate GenQu on both simulators and IBM-Q quantum computers (mainly Rome). We compare our results with Tensorflow Quantum (TFQ) and traditional convolutional neural networks with different numbers of parameters. In the rest of the evaluation, we denote CNN - xP to be classical neural networks with x parameters and QNN - xP is quantum based neural networks with x parameters.


\subsection{The Kernelized Classification}
As a proof of a single qubit natural machine learning, we employ the encoding of a circles dataset illustrated in Figure \ref{fig:circles_classical} onto one qubit through the radial kernel method. A single qubit has data points encoded as the vector distance from the center of the circle in Figure \ref{fig:circles_classical}. The qubit is then measured, and the $P(|\phi\rangle) = |0\rangle$ is equivalent to $P(Class = 1)$. Through doing so, we attain $100\%$ accuracy on separating the non-linearly separable data set, whilst maintaining both dimensions of information. Furthermore, when our experiment is run on {\bf IBM-Q's Quantum Computer Rome} 100\% accuracy is attained, thereby providing us with confidence that our model architecture works both on simulators and real quantum computers. This approach, although not novel, is done to illustrate that certain problems can be tackled very efficiently with qubits and how the solution can be successfully run on real quantum computers.

\subsection{Quantum Deep Learning}

\begin{figure}
    \centering
    \includegraphics[width=1\linewidth]{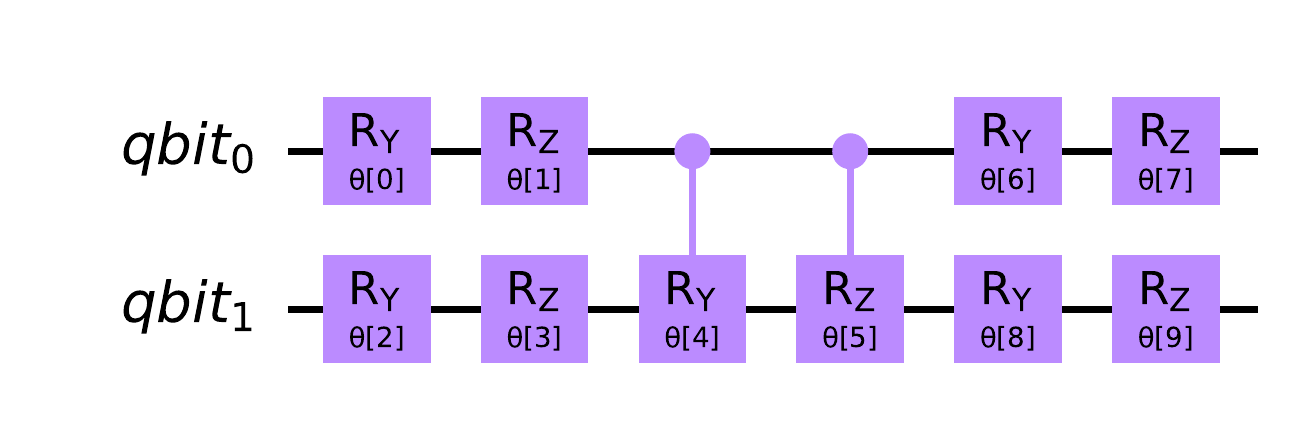}
    \caption{Quantum Deep Learning Circuit}
    \label{fig:qdl_circuit}
\end{figure}

\begin{figure*}
    \centering
        \begin{subfigure}{0.28\linewidth}
            \centering
            \includegraphics[width=1\linewidth]{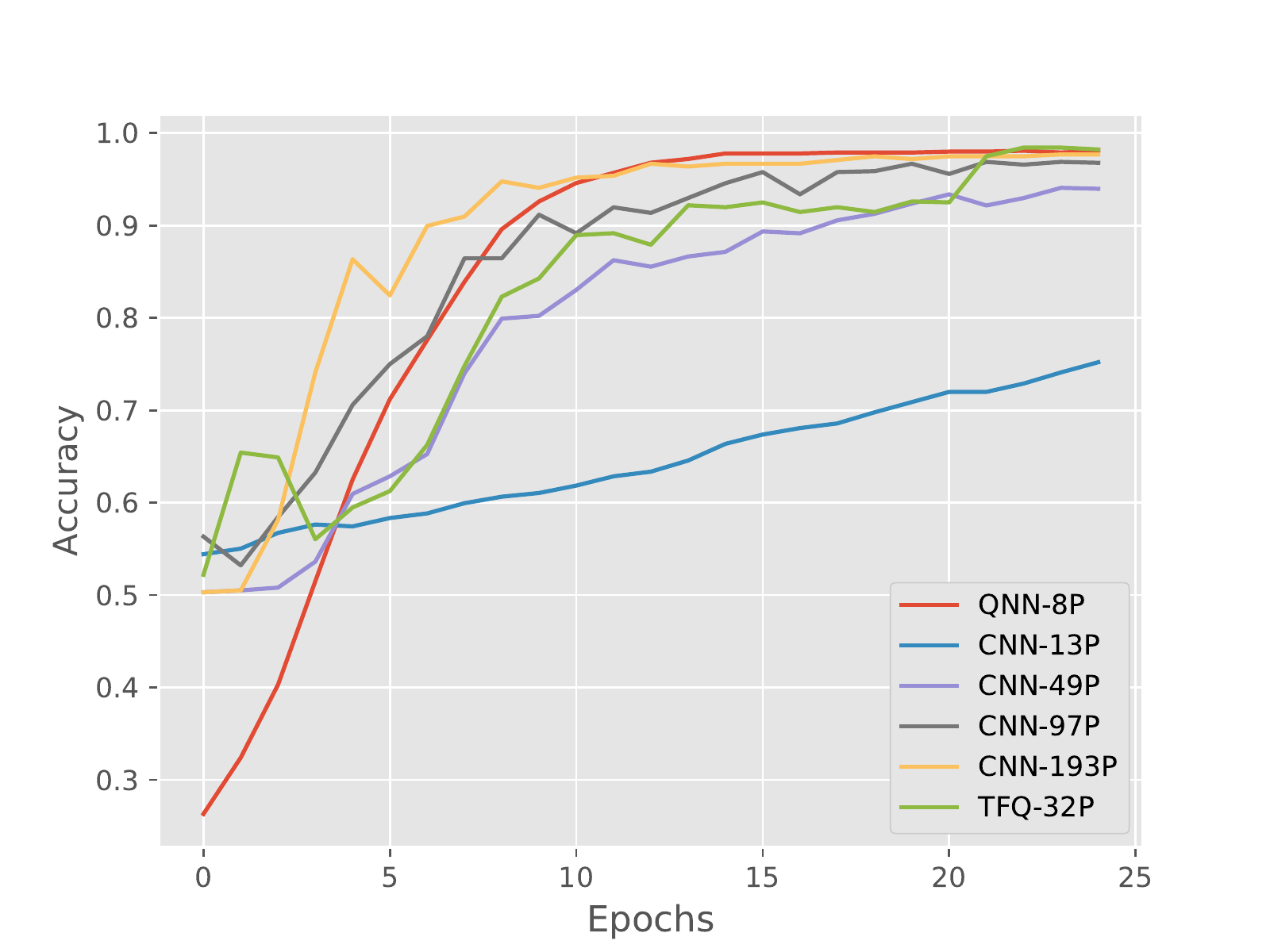}
            \caption{9 vs 6}
    \end{subfigure}
    \begin{subfigure}{0.28\linewidth}
            \centering
            \includegraphics[width=1\linewidth]{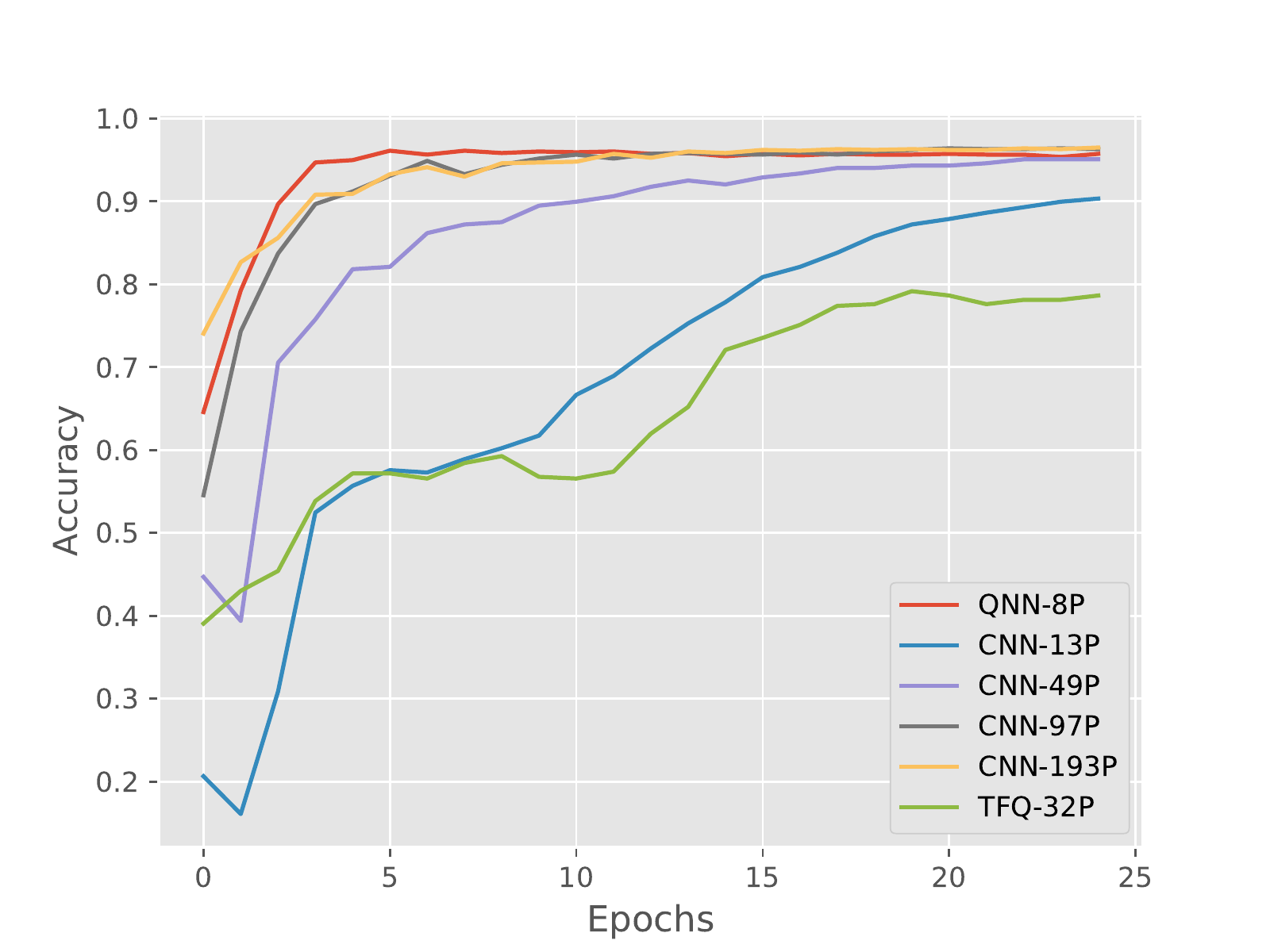}
            \caption{3 vs 1}
    \end{subfigure}
    \begin{subfigure}{0.26\linewidth}
        \centering
        \includegraphics[width=1\linewidth]{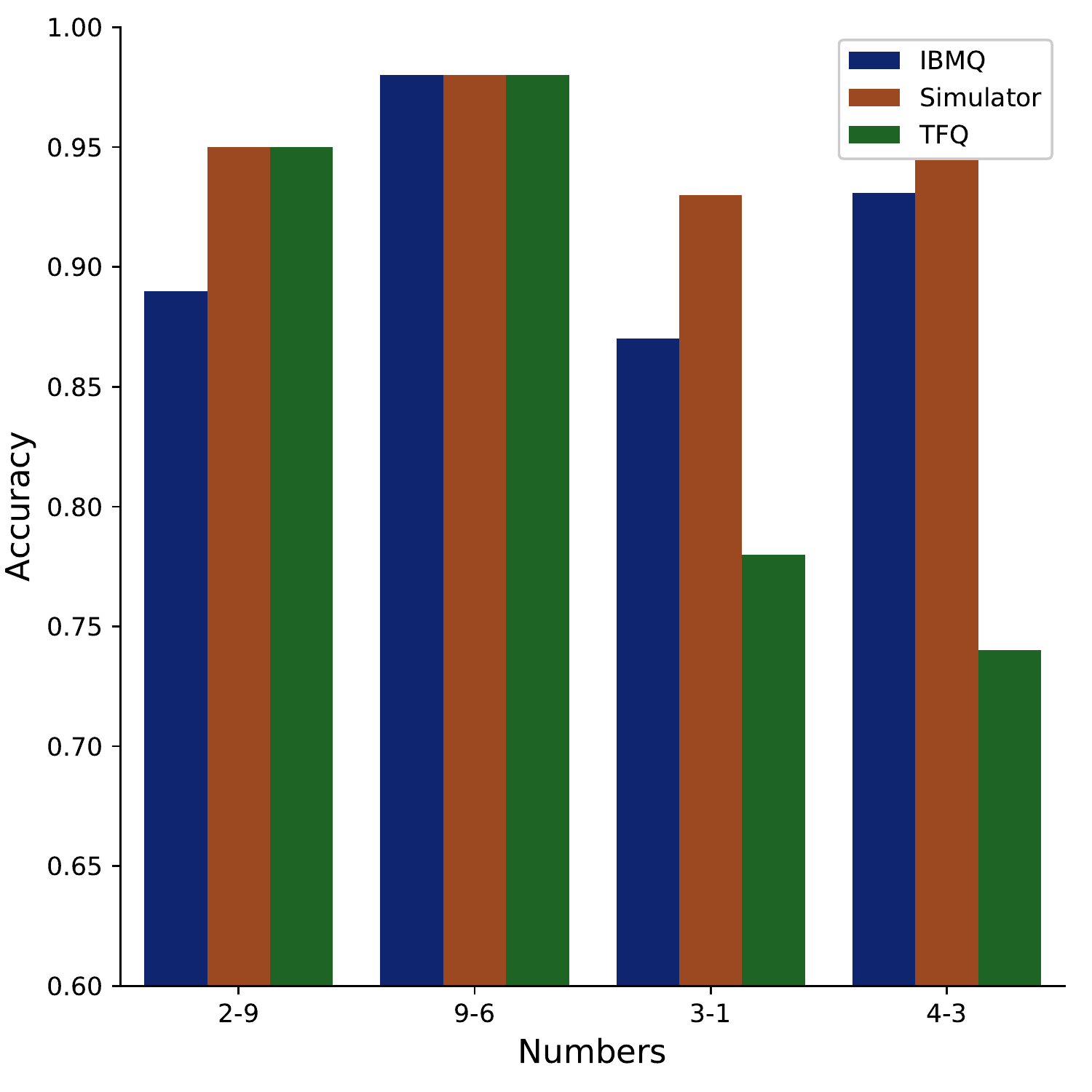}
        \caption{IBM-Q vs Simulations}
    \end{subfigure}
    \caption{QNN (simulation), QNN (IBM-Q Rome), TFQ and CNN Results}
    \label{fig:main_dnn_results}
\end{figure*}

To evaluate the learning potential of quantum deep learning, and its ability to use fewer data-channels than its classical counterpart, we make use of the MNIST data set. The MNIST data set is an image data set comprised of gray scale hand-drawn digits of resolution $28\text{ by }28$. It is infeasible to represent these images on current near-term quantum devices, and hence we make use of PCA~\citep{hoffmann2007kernel} to reduce dimensionality from $784$ to $4$. In this case, we only need to make use of 2 qubits to feed our data to our quantum deep neural network. We provide the deep learning circuit visualized in Figure \ref{fig:qdl_circuit}.
As can be seen in the circuit, there is a total of 8 parameters. This network is comprised of one single qubit unitary layer (Parameters 0 through 3), one entanglement layer which accomplishes data pooling onto one qubit (Parameters 4 and 5), and finally a single qubit unitary on the final output qubit (Parameters 6 and 7). We compare our architecture to classical deep learning architectures and compare parameter counts when using the same gradient descent approach (Adam Optimizer), same epochs, and same data set. The quantum network is trained to perform binary classification of two numbers from the MNIST dataset, where the classification is measured by the Qubit $(0,1)$ in Figure \ref{fig:qdl_circuit}. As for the classical networks, we make use of a network comprised of a middle layer of tensor size $2,8,16\text{ and }32$. The comparative training results are visualized in Figure \ref{fig:main_dnn_results}.
In Figure \ref{fig:main_dnn_results}(a), the numbers 9 and 6 are used to train the data set. The quantum network outperforms all other comparative solutions, with close to equivocal performance of a 193 parameter deep neural network, thereby attaining a 95.86\% parameter count reduction, and converging 33.33\% faster than said network. However, in the case of 9 and 6, there is a less significant difference between parameter counts than what is observed in other cases, such as 3 and 1. In Figure \ref{fig:main_dnn_results}(b), we observe how there is substantial learning ability to be gained from increasing the classical parameter count. However, similarly in this case, the 8 parameter QNN's performance is matched by the 97 parameter CNN, a 91.76\% reduction in parameters. We also compare our model to the Tensorflow Quantum (TFQ) MNIST classification example on the same number pairs\cite{broughton2020tensorflow}. We illustrate that our network outperforms Tensorflow Quantums MNIST classification task in Figures \ref{fig:main_dnn_results} (a) and (b). This illustrates the significant learning potential of quantum networks and specifically the architecture used in this paper. These architectures are able to, in certain cases, reduce parameter counts significantly with no sacrifice to performance. Furthermore, in our case we have encoded two dimensions of data per qubit. Feeding 4 dimensions of data to a deep neural network through 2 {\em neurons} is impractical, and is a further example of how powerful qubits are in deep learning.

We validate our results by running similar experiments on a real quantum computer using the IBM-Q platform, comparing the accuracy's attained on a simulator to that of a on a quantum computer. These results are visualized in Figure \ref{fig:main_dnn_results}(c). As can be seen, for numbers \textbf{4-3} and \textbf{9-6}, actual quantum computer performance was extremely similar to that of the simulator, with a difference of less than 5\%, and 9-6 having a measured difference of 0.2\%. However, in the case \textbf{3-1} we observe more significant differences between actual Quantum Computing implementation. The largest difference between simulators and actual quantum computers was 7.25\% on the \textbf{3-1} dataset, which is due to the noise on the quantum computer that depends on the computer itself and the rotated workloads on it (random factors). We further validate our results by comparing them to the TFQ MNIST classification task, and show how in some cases our architecture and network attaining significantly higher accuracy. Specifically, in the case of \textbf{4-3} with a 20\% improvement and \textbf{3-1} with a 22\% improvement. However, in certain cases our network attained the same accuracy as TFQ, in the cases of \textbf{2-9} and \textbf{9-6}.

\subsection{Quantum Generative Nature}
\begin{figure*}
    \centering
    \begin{subfigure}{0.26\linewidth}
        \centering
        \includegraphics[width=1\linewidth]{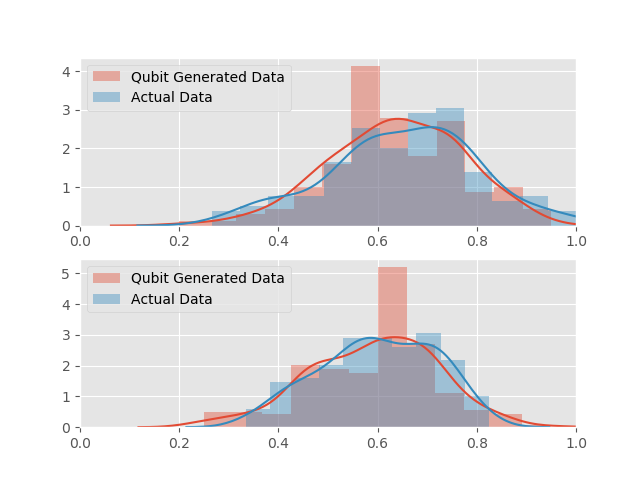}
        \caption{Epoch 25 - QNN - 2P}
    \end{subfigure}
    \begin{subfigure}{0.28\linewidth}
        \centering
        \includegraphics[width=1\linewidth]{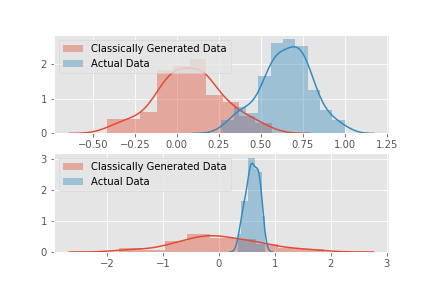}
        \caption{Epoch 25 - CNN - 196P}
    \end{subfigure}
        \begin{subfigure}{0.28\linewidth}
        \centering
        \includegraphics[width=1\linewidth]{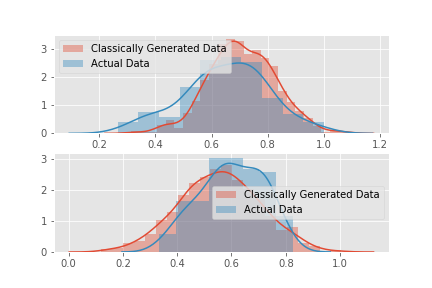}
        \caption{Epoch 25 - CNN - 2144P}
    \end{subfigure}
     \caption{Single qubit generative model to learn the distribution of PCA MNIST digit 0}
    \label{fig:generative_results}
\end{figure*}

\begin{figure*}
    \centering
    \begin{subfigure}{0.21\linewidth}
        \centering
        \includegraphics[width=1\linewidth]{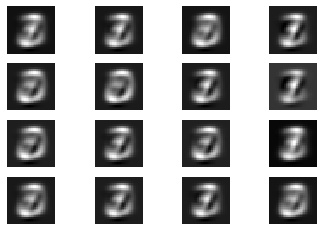}
        \caption{Epoch 0}
    \end{subfigure}
    \hspace{0.05\linewidth}
    \begin{subfigure}{0.21\linewidth}
        \centering
        \includegraphics[width=1\linewidth]{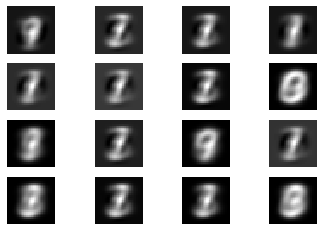}
        \caption{Epoch 25}
    \end{subfigure}
    \hspace{0.05\linewidth}
        \begin{subfigure}{0.21\linewidth}
        \centering
        \includegraphics[width=1\linewidth]{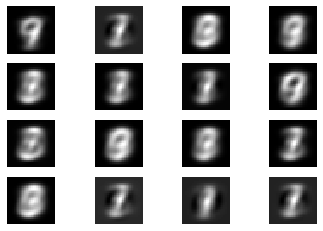}
        \caption{Epoch 50}
    \end{subfigure}
    \caption{2-Dimension Images Generated by GenQu}
    \label{fig:images}
\end{figure*}

Another point of interest is how powerful qubits are in representing data sets. This has significant implications in Generative Adversarial Networks (GANs) and loading data sets. We illustrate this potential by minimizing the distance between a single qubit`s quantum state and the MNIST data set PCA'ed to 2 dimensions and of class 0. We illustrate in Figure \ref{fig:generative_results} how a single qubit, visualized by the blue shading, using only 2 parameters (an RY and RZ gate in series on one qubit), can completely mimic the data it was fed (2 dimensions). If sampled, the qubit will generate all samples it was fed as well as generate new unique samples similar to that of which it was fed. We make use of the architecture of a Generative Adversarial Network \cite{goodfellow2014generative} to compare this to a classical neural network, and observe how poorly the classical counterpart performs. When given 9800\% more parameters (2 vs 196), as visualized in Figure \ref{fig:generative_results}(b), the network was still unable to mimic the data fed to the network. The classical network was able to converge when provided with $2144$ parameters, as visualized in Figure \ref{fig:generative_results}(c). Furthermore, we illustrate the generative potential by sampling from the aforementioned quantum circuits with a shot count of 15, performing reconstructive PCA on the data and plotting the images. As seen in the images the qubit is able to learn from almost nothing in \ref{fig:images}(a) to reasonable results after 50 epochs in \ref{fig:images}(c). This further illustrates the significant machine learning potential and parameter reduction potential of quantum machine learning. 

\section{Conclusion}

This paper proposes GenQu, a hybrid and general-purpose quantum framework for learning classical data. It extends the Tensorflow Quantum framework such that the applications can be executed on public-available quantum computers (IBM-Q Platform). GenQU demonstrates the significant expressibility of qubits, and their extensive applications in machine and deep learning. 
Based on GenQu, we propose three different learning models that make use of a low-qubit count in near-term quantum computers.
In the model for kernelized classification, GenQu is able to to encode the circle dataset onto a single qubit achieve 100\% accuracy in the experiment on a real quantum computer.
With quantum deep learning model in GenQu, when encoding two dimensions of data per one qubit, it is able to show reductions in parameters equivalent to 95.86\%, whilst still attaining a similar accuracy or better than that of classical deep learning models. 
Finally with respect to qubits learning potential, a single qubit generative model is proposed. It is able to completely learn to generate 2 dimensions of data PCA'ed from the MNIST's 0 class. As the future work, we plan to fully integrate GenQu with Tensorflow Quantum and add more build-in features, such as amplitude encoding, customized evaluation functions. 




\section*{Acknowledgment}
This material is based upon work supported by the U.S. Department of Energy, Office of Science, National Quantum Information Science Research Centers, Co-design Center for Quantum Advantage ($C^2QA$) under contract number DE-SC0012704. 


\bibliography{main.bib}
\bibliographystyle{IEEEtran}

\appendix

\section{Appendix: Experiments on IBM-Q quantum computers}

\begin{figure}[ht]
   \centering
         \begin{minipage}[t]{0.45\linewidth}
\centering
        \includegraphics[width=0.95\linewidth]{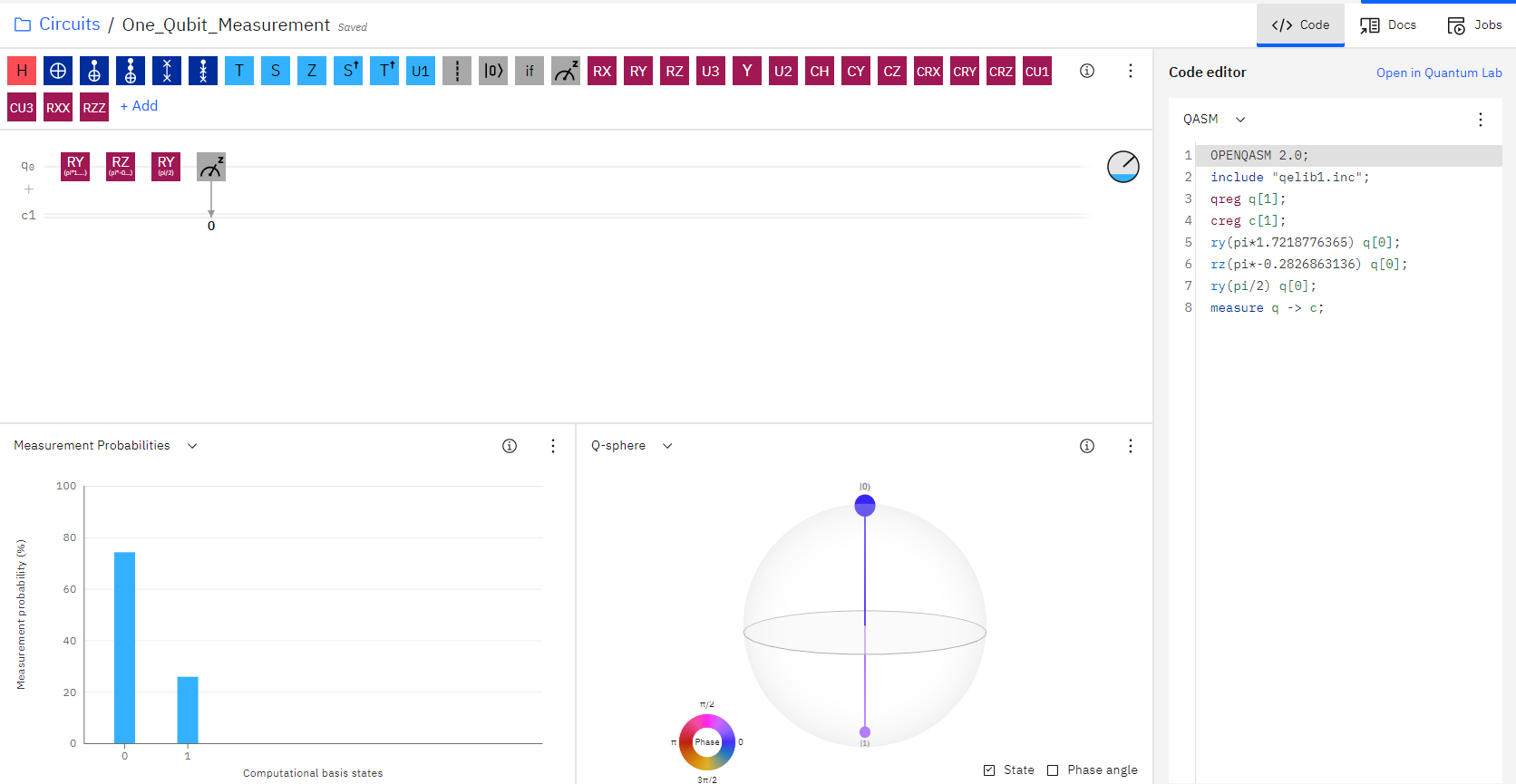}
    \caption{Experiment setup for Kernerlized classification}
    \label{fig:ibm-circuit-kernel}
      \end{minipage} 
      ~
      \begin{minipage}[t]{0.45\linewidth}
\centering
      \includegraphics[width=0.95\linewidth]{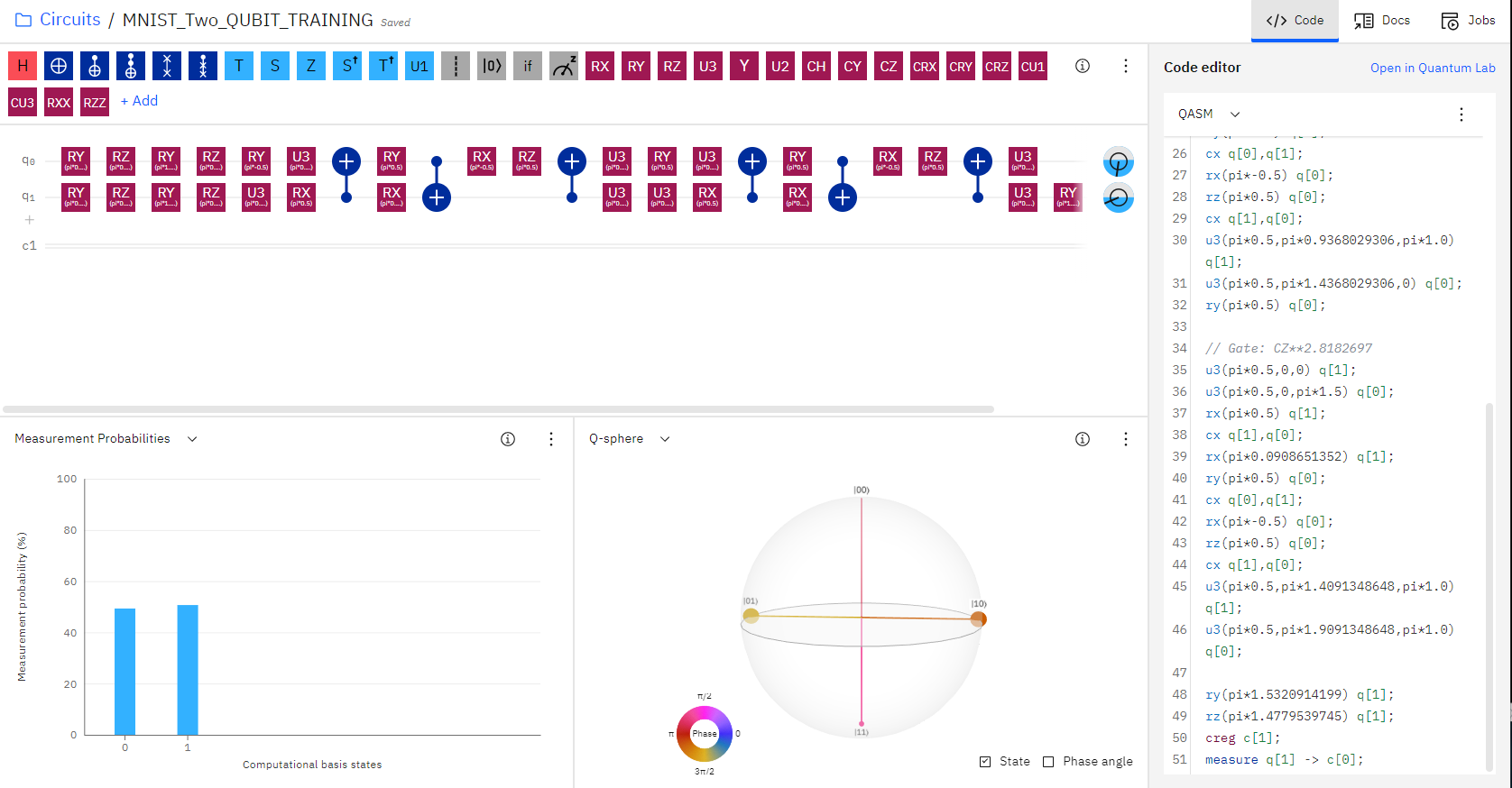}
    \caption{Quantum circuit on MNIST dataset (T1)}
    \label{fig:ibm-circuit-t1}
      \end{minipage} %
\end{figure}

To justify our proposed model, 
we conducted experiments on multiple  IBM-Q sites including Vigo, Ourense, Rome, Bogota and Valencia. 
Figure~\ref{fig:ibm-circuit-kernel} is the setup for Kernelized classification on the circle dataset. We ran this circuit with 40 shots (repetition) 20 times on different locations and achieved accuracy of 100\% . 

\begin{figure}[ht]
   \centering
         \begin{minipage}[t]{0.45\linewidth}
\centering
       \includegraphics[width=0.95\linewidth]{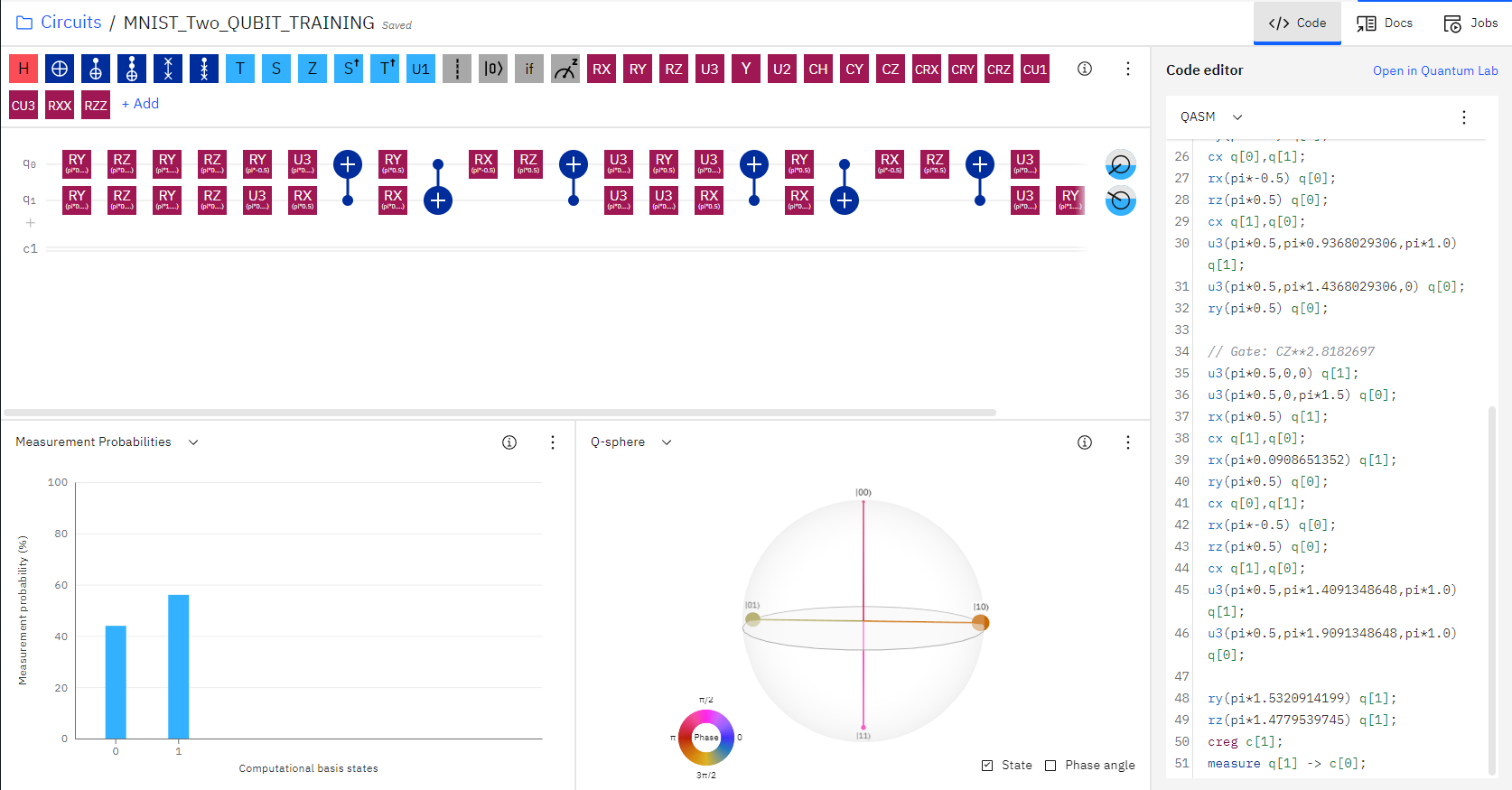}
    \caption{Quantum circuit on MNIST dataset (T2)}
    \label{fig:ibm-circuit-t2}
      \end{minipage} 
      ~
      \begin{minipage}[t]{0.45\linewidth}
\centering
       \includegraphics[width=0.95\linewidth]{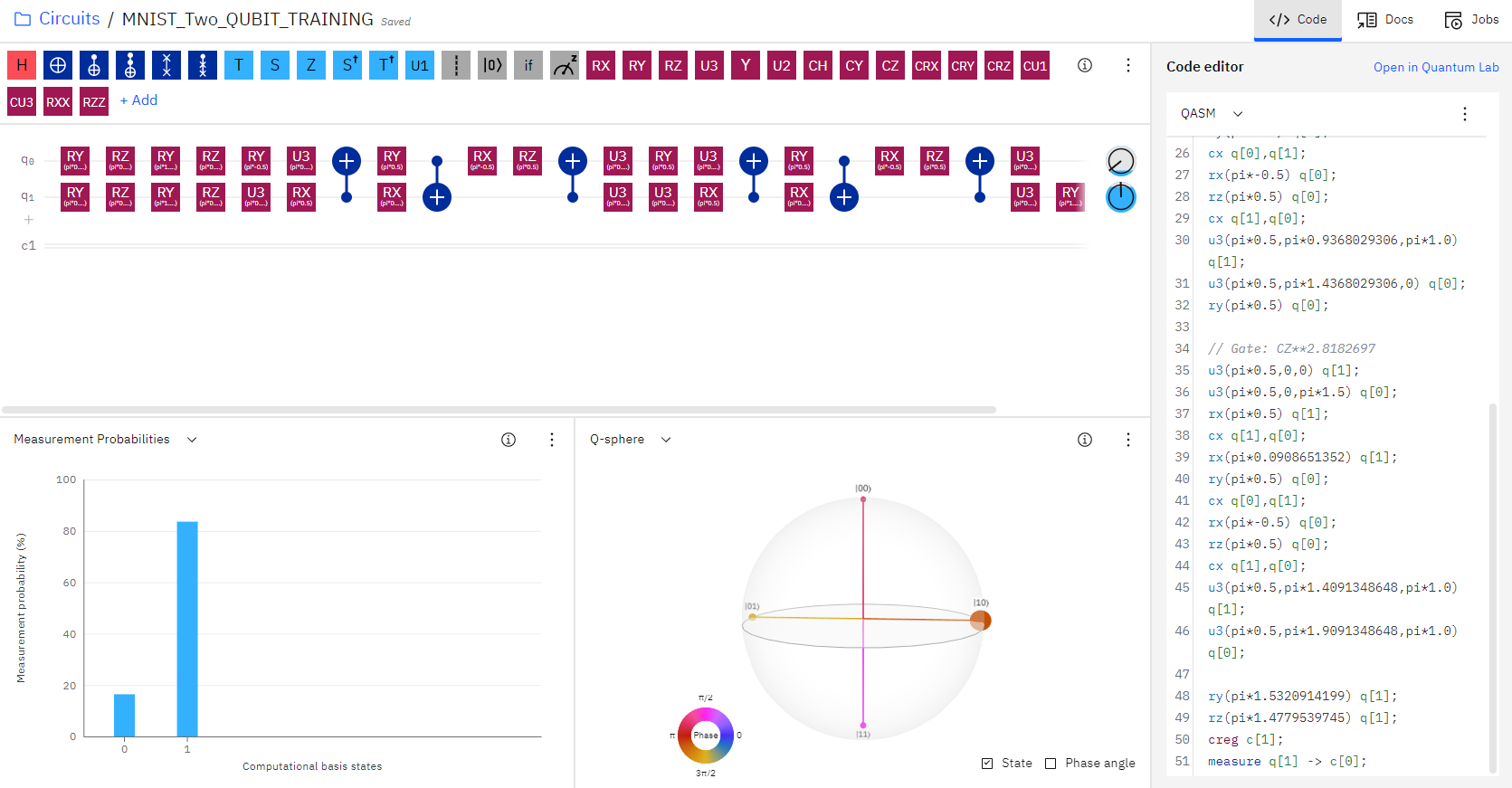}
    \caption{Quantum circuit on MNIST dataset (T3)}
    \label{fig:ibm-circuit-t3}
      \end{minipage} %
\end{figure}

In PCA MNIST dataset, we train the model on 913 samples of two specific digits (9 and 6) on different IBM-Q locations to justify our founding. 
Figure~\ref{fig:ibm-circuit-t1}, \ref{fig:ibm-circuit-t2} and \ref{fig:ibm-circuit-t3} present a sample of each iteration over actual IBM-Q machine which ran for 913 times to get accuracy of 95\% on average. 

\end{document}